# Ultra-Low Noise Amplifier Design for Magnetic Resonance Imaging systems

Mustafa A. Kabel – orcid.org/0000-0002-2976-4333

*DTU Electro – Technical University of Denmark*
*Electromagnetic Systems Group*
*Ørsteds Plads 348 – 2800 Kgs. Lyngby − Denmark*

Dated: May, 2017

*Abstract* − **This paper demonstrates designing and developing of an Ultra-Low Noise Amplifier which should potentially increase the sensitivity of the existing Magnetic Resonance Imaging (MRI) systems. The Design of the LNA is fabricated and characterized including matching and input high power protection circuits. The estimate improvement of SNR of the LNA in comparison to room temperature operation is taken here. The Cascode amplifier topology is chosen to be investigated for high performance Low Noise amplifier design and for the fabrication. The fabricated PCB layout of the Cascode LNA is tested by using measurement instruments Spectrum Analyzer and Vector Network analyzer. The measurements of fabricated PCB layout of the Cascode LNA at room temperature had the following performance, the operation frequency is 32 MHz, the noise figure is 0.45 dB at source impedance 50 Ω, the gain is 11.6 dB, the output return loss is 21.1 dB, and the input return loss 0.12 dB and it is unconditionally stable for up to 6 GHz band. The goal of the research is achieved where the Cascode LNA had improvement of SNR.**

## I. Introduction

An LNA combines a low noise figure, reasonable gain, linearity and stability without oscillation over entire useful frequency range. The LNA function, play an important role in the receiver designs. Its main function is to amplify extremely low signals without adding noise, thus preserving the required Signal-to-Noise Ratio (SNR) of the system at extremely low power levels. Our LNA is designed for MRI scanner and since the MRI signal is very low, it needs to be amplified to a level that is sufficient for transmission along long cables and further processing [1-4]. Noise is inevitably added to the signal during this amplification so the signal-to-noise ratio (=SNR) of the MRI signal is decreased [5] and this will decrease the diagnostic accuracy of MRI which is vitally important in medical applications. The noise contribution during the amplification is defined by noise factor F, being the ratio of input SNR to output SNR. According to the Friis formula [6], if the preamplifier (amplifier in the first stage) has gain, the noise from the following next amplifiers can be neglected and the preamplifier basically dominates the total noise factor of the whole receiving chain. The researching of exciting amplifier circuit topologies and suitable semiconductor technologies is conducted. One of the most important factors to achieve low SNR at LNA is choosing the right transistor to be built as LNA. PHEMT ATF54143 transistor is chosen for our Cascode LNA design which it contributes very low noise figure at its optimal source impedance $Z_{opt}$.

The Cascode amplifier topology is designed in this paper as Ultra LNA and the fabrication of the Cascode LNA is constructed on PCB FR4-board. Additionally, The main characteristic of Cascode amplifier it has advantage of higher input-output isolation. This means there is no direct coupling from the output to input and then reduce the power that will be wasted in part of amplifier circuit at amplifying the input signal. The Low Noise Amplifier (LNA) operates in Class A and the general topology of the LNA consists of three stages [7], the input matching circuit, the amplifier itself (transistor) and the output matching circuit as shown in figure 1.The input and output matching circuit are one of the important requirements for achieving the desired performance of LNA. In amplifier design, we need to ensure that input and output matching network must meet the criteria required for low noise, stability, small signal gain and linearity [8]. Other requirements that they have been taken into account in design of the Cascode LNA, the loss of components and the loss at RF paths (Traces that connect the components on PCB). These losses contribute to increasing of noise figure factor at the LNA. Therefore it has to design the LNA circuit on PCB which has specific components types and short traces at the input of the LNA which offer minimum losses [9].

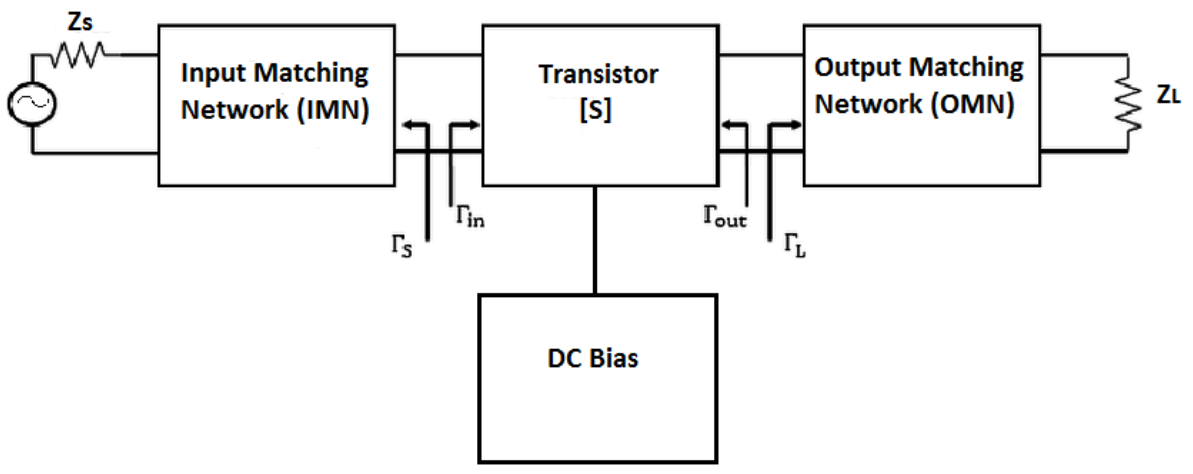

Figure 1. Block Diagram of typical LNA

The Cascode LNA should be designed to have protection at input high power during the transmitting period of MRI system. Therefore it used a technique of using 2 shunt Pin diodes at the input of the Cascode LNA to limit the input high RF pulse power. The Cascode LNA should also be designed to interface low impedance sources and it should be unconditionally stable. This paper presents researching by first introducing the relevant theatrical background on description of circuit design of the Cascode LNA and its performance and results of measurements of the fabricated Cascode LNA.

## II. Research Methods

The design methodology of Cascode LNA found the DC operating point for minimum noise figure at the transistor ATF54143 and the design of proper DC biasing network for the LNA and the analysis of the LNA stability, input & output matching network and Linearity of the designed LNA. The designed Cascode LNA Circuit transferred into PCB Layout. Then the Cascode LNA is simulated by using EM/Circuit Co-Simulation capability in Advance Design System ADS.

### *A.* Description of the Cascode LNA Circuit

The circuit of the developed Cascode LNA is shown in Fig. 2. It is a single stage Cascoded LNA amplifier. The circuit of the Cascode LNA consists of a common-source input stage loaded by the source of a common-gate output stage [10]. The input stage of the Cascode LNA is driven by the signal RF in. Then this input stage drives the output stage, with output signal RF out. The DC bias operating point is chosen from datasheet to exhibit stable thermal performance and minimum noise figure for Drain current $I_{DS}$ = 18 mA at the transistor ATF54143. The bias circuit consists of resistive network R7, R8, and R9 which is voltage dividers that provided the specific DC voltages, 530 mV at the Gate of transistors Q1 and 570 mV at the Gate of transistors Q2. The DC supply voltage VDD is chosen to be 5 volts because the Cascode amplifier topology needs more DC supply voltage than other topologies.

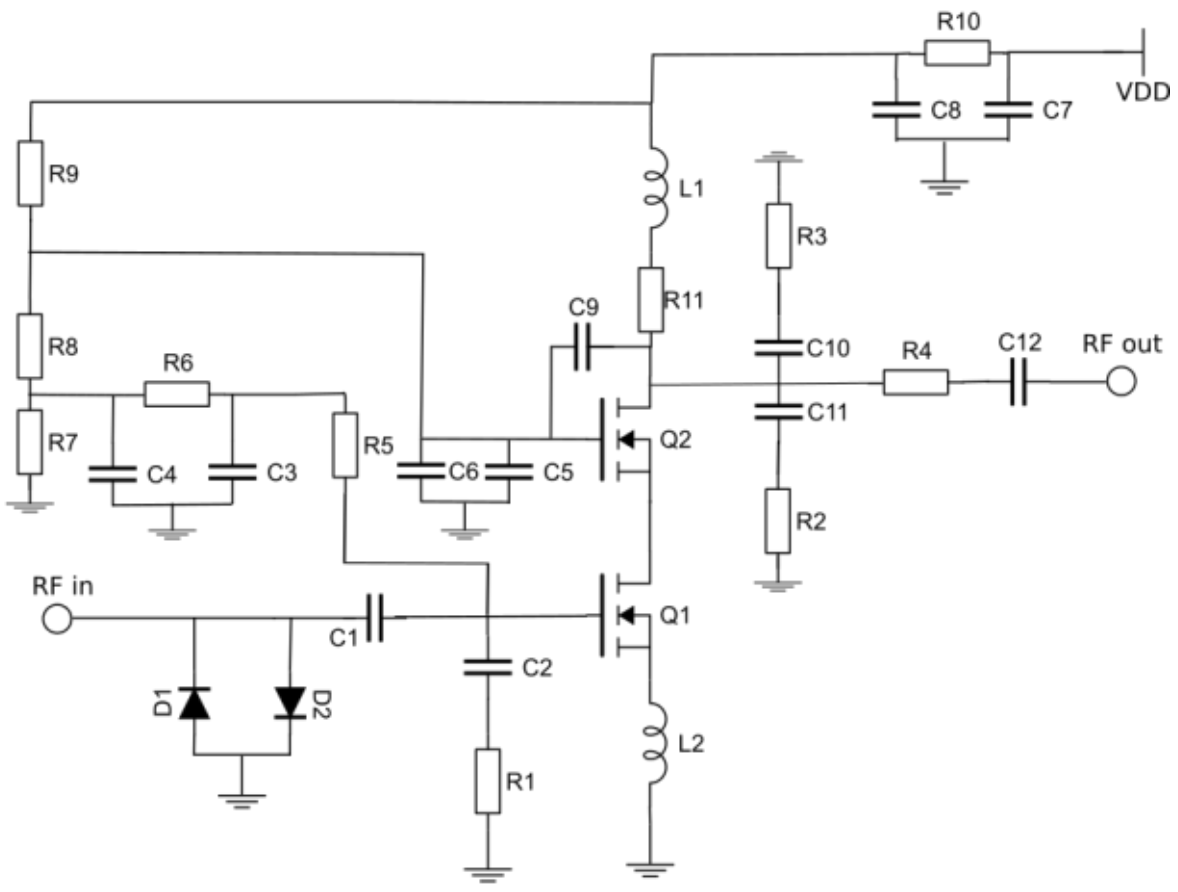

Figure 2. The Cascode LNA Circuit

The capacitors C3, C4, C7, and C8 are bypass capacitors that shorts AC signals to ground, so any AC noise that may be present on a DC supply voltage or any RF-energy from the transistors is removed, producing a much cleaner and pure DC bias voltage to the Gate of the transistors. The capacitors C5 and C6 are also bypass capacitors that provide path for RF signal to be shunted to ground. The resistors R6 and R10 provide important low frequency termination for the device and it improves stability at low frequency [11]. The resistor R5 provide DC current limiting for the Gate of the device Q1 and this is important when the device is driven to $P_{SAT}$ [11]. The resistor R9 also used for blocking RF signal to isolate RF path (act as RF choke) while passing DC bias voltage to the Gate of the device. The inductor L1 is RF choke. The capacitors C1 and C12 are DC Block capacitors and its values are calculated to show reactance equal to 0.5 ohm on RF path at 32 MHz.
It is applied the shunt resistors R1 and R2 to the input and output of the Cascode LNA, these components stabilized the Cascode LNA at High frequency by adding loss to the input impedance and output impedance of the Cascode LNA [12-13]. In order to achieve great stability at the Cascode LNA, the capacitance parallel feedback technique is applied and the capacitor C9 increased the stability at High frequency [14]. Additionally, the stability at higher frequencies is increased more in the circuit by adding the inductor L2 which is inductive source degeneration [7, 15]. The stability of the Cascode LNA at Low frequency is improved by the applied shunt resistor R3 to the output of the Cascode LNA [12-13]. The resistive loss R11 is added to the circuit design in order to increase more the stability at low frequency [16]. The capacitors C2 and C11 values are used to get impact of the shunt resistance of the stability on High frequency Band while the capacitor C10 value is used to get impact of the shunt resistance on Low frequency Band. The resistor R4 is output resistive loading and it is added to the output of the Cascode LNA for optimizing the stability and for the matching purpose.

The Cascode LNA designed to have built-in Pin diodes at the input port to handle high input RF pulse power level during the transmitting period in MRI system. The protection technique is using 2 shunt Pin diodes to ground at the input of the Cascode LNA. The diode technique is showed above in Fig. 2 and they are diodes D1 and D2 which are the High Power Pin Diode MA4P1250NM-1072T in the Cascode LNA circuit.

*B.* PCB Layout of the Cascode LNA

The Cascode LNA circuit is transferred into PCB Layout as it showed in Fig. 3. The PCB board was FR4 and it is designed as microstrip board. There are benefits of choosing Microstrip technology. The bottom layer of microstrip board is completely ground plane, this bottom layer provided short RF path for ground nodes of the circuit on top layer.

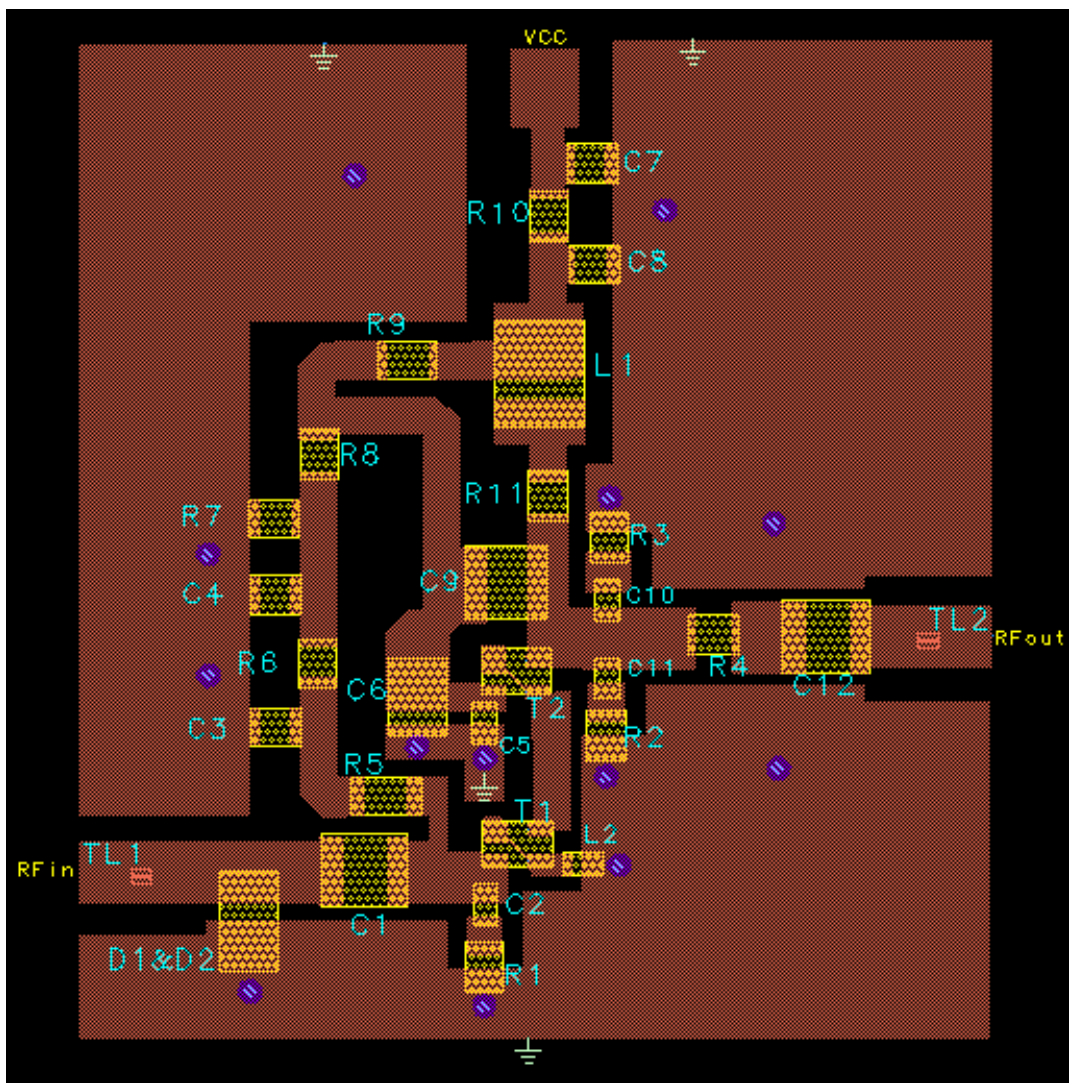

Figure 3. PCB Layout of the Cascode LNA

The reducing of RF path length implied less parasitic inductances which are desired in RF/microwave layout design [17]. The traces at input and output RF paths of the Cascode LNA have been designed as microstrip transmission lines TL1 and TL2 which have characteristics impedance $Z_o$ = 50 Ω at 32 MHz. This can improve input and output return loss on the Cascode LNA [9]. The package footprints of the components are chosen to be SMD in order to achieve compact PCB Layout design. It also applied non-magnetic lumped components for DC-blocks capacitors and for the parallel feedback stability capacitor in order to achieve minimal magnetic and less parasitic components in the PCB board which is desirable in components of MRI system.

On the top layer of the PCB Layout, there is ground plane which is composed of three areas on the top layer. This ground plane is at least 0.25 mm away from the components and traces. These areas provide short path for circuit to ground and minimizing current loop. The bypassing capacitors are connected to its own ground by using these areas for providing low impedance path for bypassing. The bottom (flat copper) layer of the microstrip board allowed for solid ground plane under the Cascode LNA circuit and the ground nodes of the circuit on top layer are connected to the bottom layer over circular via holes to achieve high performance at the Cascode LNA. Parasitic inductances in RF paths affect the performance of the circuit design [17], therefore it avoided long traces that have parasitic inductance associated with it when the components placements have been drawn.

*C.* Stability of the Cascode LNA

The stability of the Cascode LNA is a very important factor which must not be susceptible to unwanted oscillation. The Cascode LNA is designed to be unconditionally stable at the whole 6 GHz band which is the Low Noise PHEMT transistor ATF54143 Band. It could present any load to the input or output of the Cascode LNA, including open or short circuit load. The stability of the Cascode LNA could be determined from its *S*-parameters and the load and source impedances over frequencies. A new criterion has been proposed that combines the scattering parameters in a test of unconditionally stable amplifier involving only a single parameter $\mu$, " Mu Stability Factor" [7] and it is given in equation (1).

$$\mu = \frac{1 - |S_{11}|^2}{|S_{22} - \Delta S_{11}^*| + |S_{12}\, S_{21}|} > 1 \qquad (1)$$

Where $S_{11}^*$ is a complex conjugate of $S_{11}$ and $\Delta = S_{11}\, S_{22} - S_{12}\, S_{21}$

Thus, if $\mu > 1$, the amplifier is unconditionally stable. In addition, it can be said that larger values of $\mu$ imply greater stability. The parameter $\mu$, Mu Stability Factor of the Cascode LNA is achieved as shown in Fig. 4 and it is bigger than one over frequencies up to 6 GHz at the simulation of the Cascode LNA. The simulation showed the Cascode LNA is unconditionally stable and it will not oscillate when its input and output are terminated with impedances having zero or positive resistances.

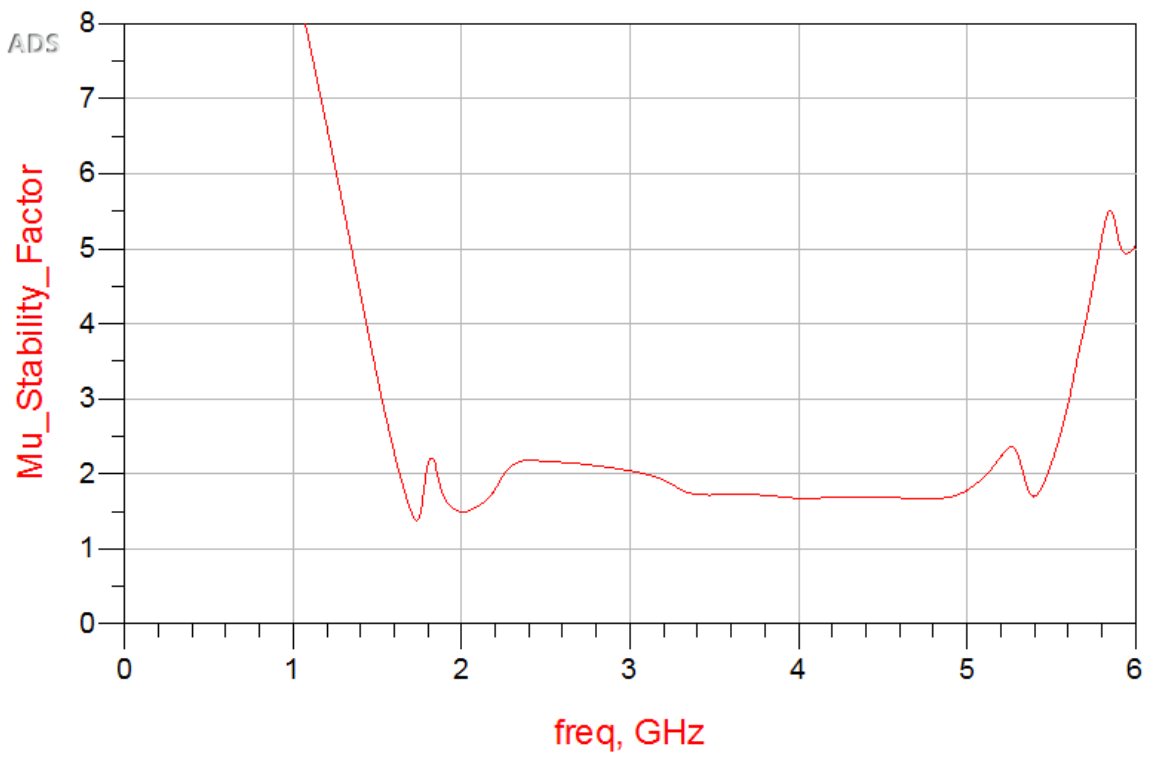

Figure 4. Simulated Mu Stability Factor of the Cascode LNA

Furthermore the Cascode LNA is unconditionally stable because its input and output impedance over frequencies are not clipped the outer edge of the Smith chart [7, 12, 16] as shown in Fig. 5. The simulation showed the input impedance of the Cascode LNA is 37.7 – j387.6 Ω and the output impedance is 45.7 – j4.1 Ω at the operation frequency 32 MHz.

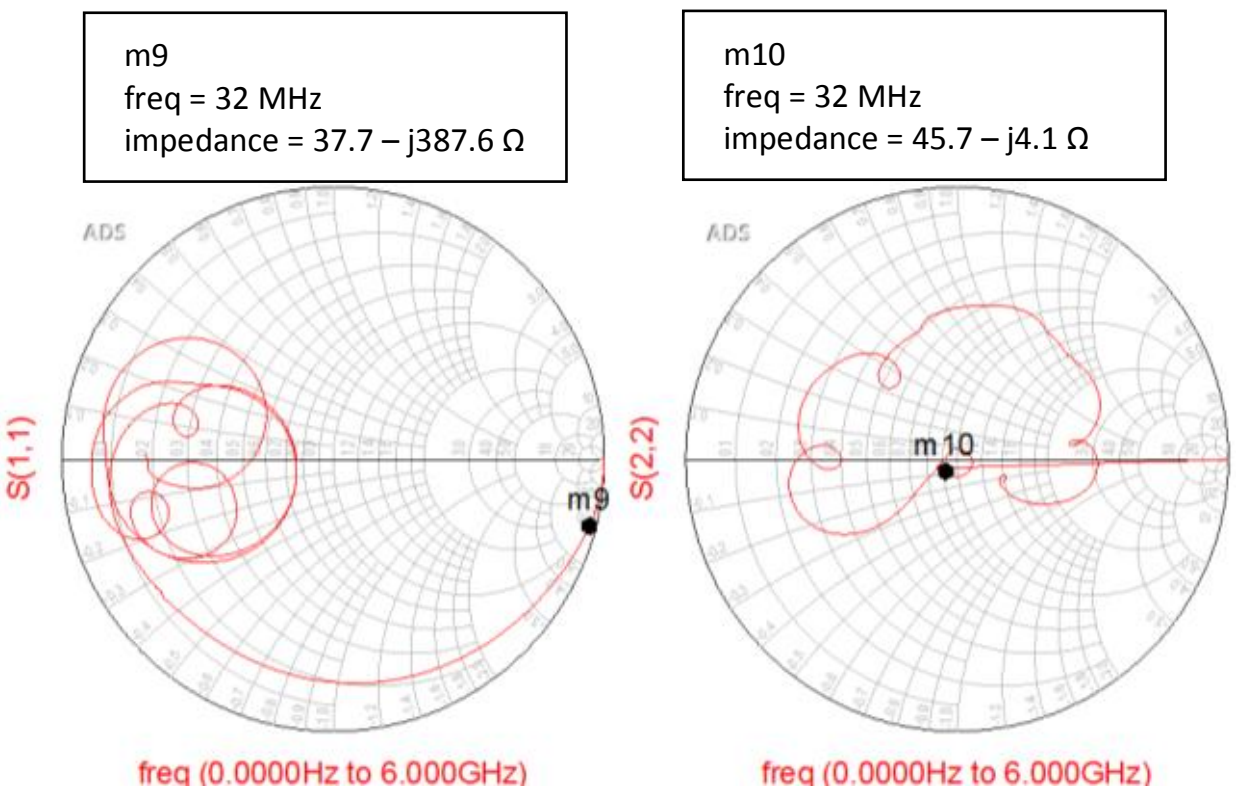

Figure 5. Simulated Input Impedance and Output Impedance of the Cascode LNA over frequencies in Smith chart

Instabilities of an amplifier are primarily caused by three phenomena: internal feedback of the transistor, external feedback around the transistor caused by external circuit, or excess gain at frequencies outside of the band of operation. It could design the Cascode LNA to be unconditionally stable up to 6 GHz Band by using the stability techniques in the circuit design.

*D.* Noise Figure and S-parameters Characteristics in the Cascode LNA

Noise figure F of an LNA depends on semiconductor devices (transistor noise parameters). As shown in equation (2), transistor noise performance is independent of load termination and is determined by its source termination and noise parameters [7-8]. The noise parameters fully describe the noise performance of a device for a specific set of conditions such as frequency, bias, and temperature.

$$F = F_{min} + \frac{4r_n \,|\, \Gamma_s - \Gamma_{opt} \,|^2}{(1 - |\,\Gamma_s|^2)\,|\,1 + \Gamma_{opt}|^2} \qquad (2)$$

Where $r_n$ is normalized equivalent noise resistance of transistor, $\Gamma_s$ is reflection coefficient of the source, $F_{min}$ is minimum noise figure of transistor and $\Gamma_{opt}$ is Gamma optimum.
Transistor noise figure F is a function of $\Gamma s$, $F_{min}$, $r_n$, and $\Gamma_{opt}$ where $F_{min}$, $r_n$, and $\Gamma_{opt}$ are known as the transistor noise parameters. As reflection coefficient of the source $\Gamma_s$ approaches $\Gamma_{opt}$, the transistor noise figure approaches its minimum. As $\Gamma_s$ departs from $\Gamma_{opt}$, the noise figure increases from its minimum value [9]. The simulation of noise figure of the Cascode LNA is shown in Fig. 6.

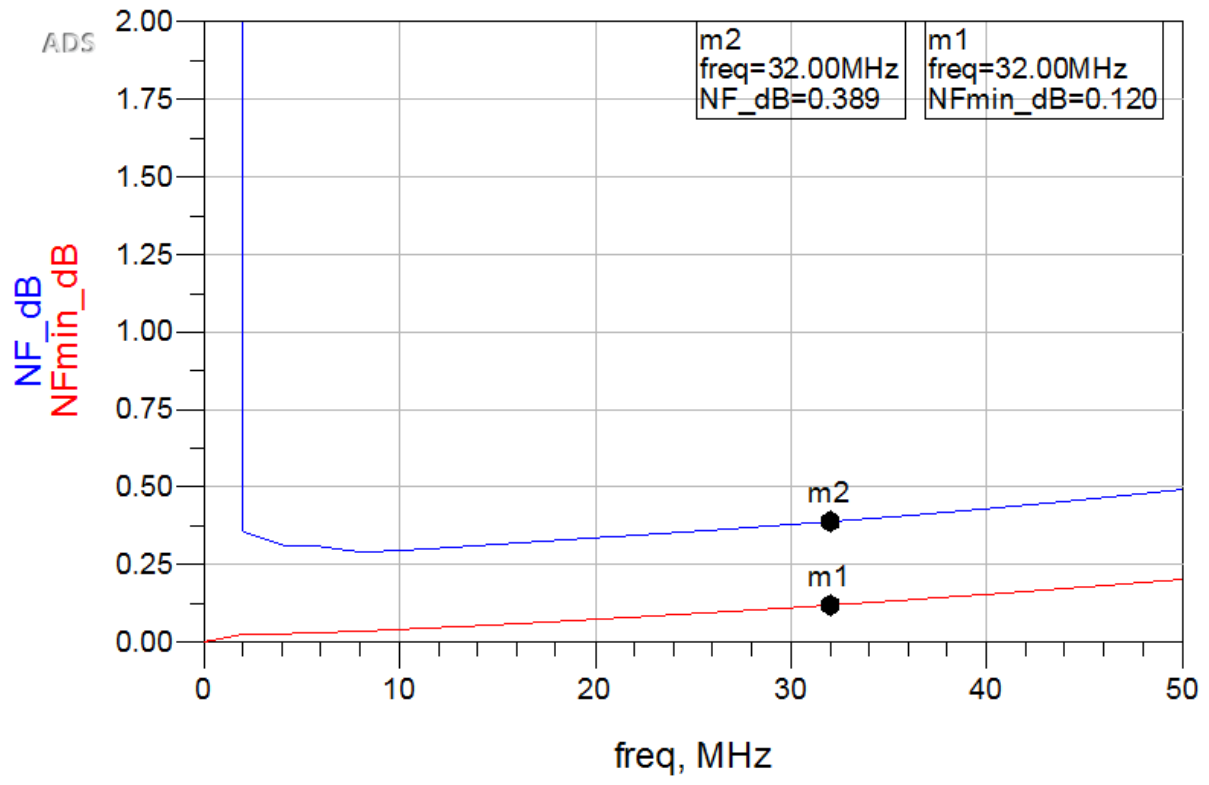

Figure 6. Simulated Noise Figure at the Cascode LNA

The magnitudes of S-parameters in dB of the Cascode LNA are achieved at the simulation as shown in Fig. 7. The gain is 13.7 dB and the output return loss is 24.1 dB. Furthermore, the input return loss is 0.21 dB and the reverse isolation (Input-Output) is 93 dB.

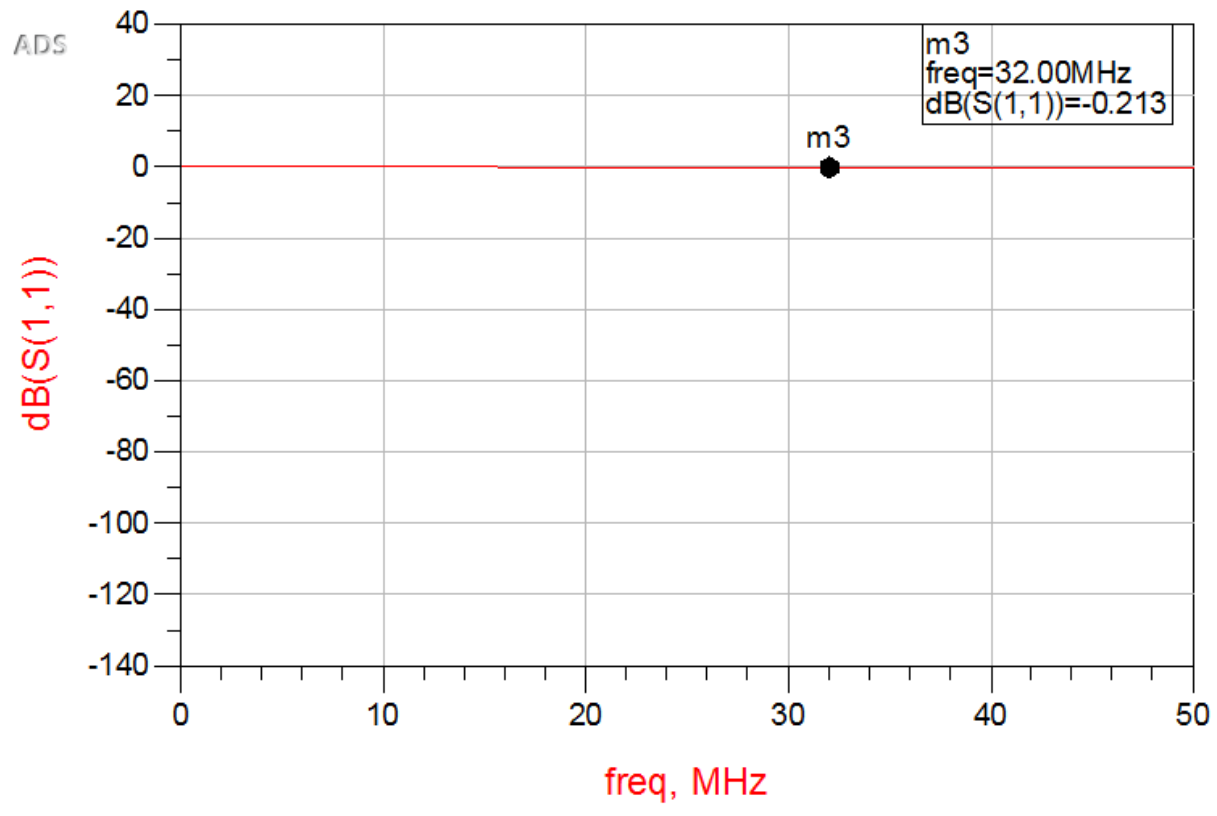

A)

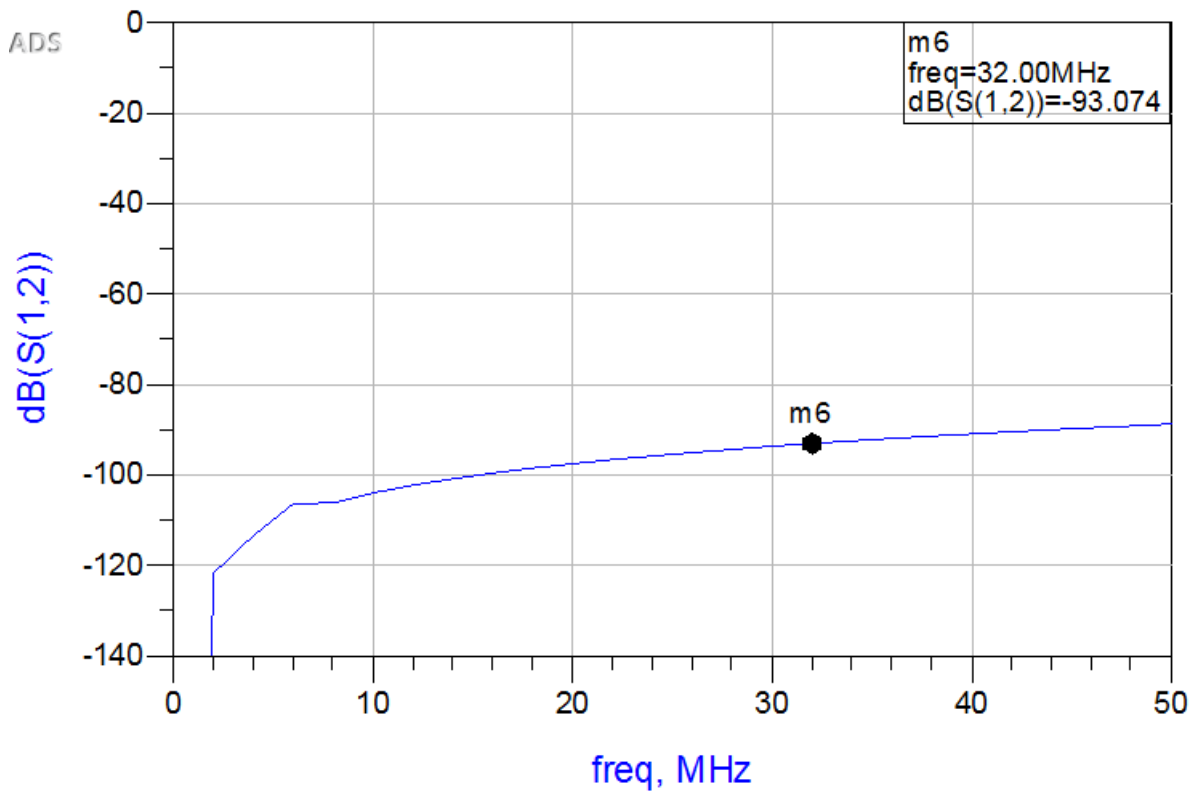

B)

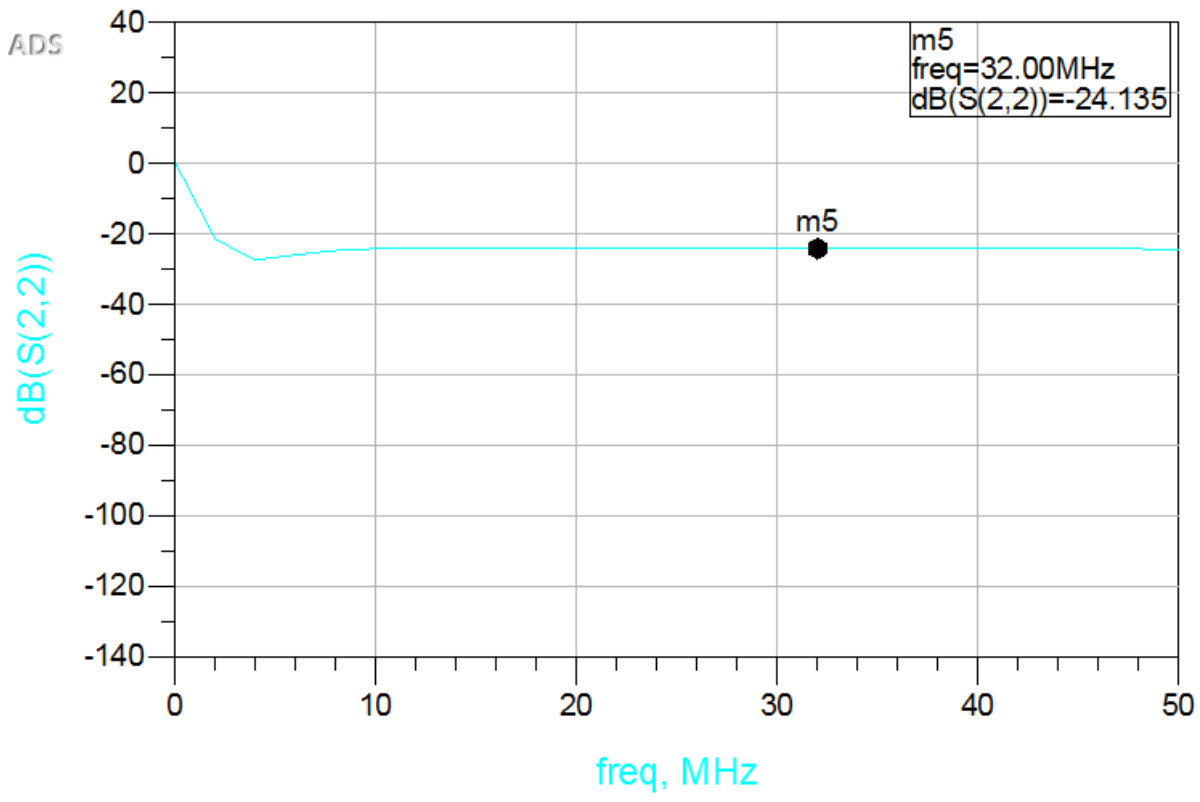

C)

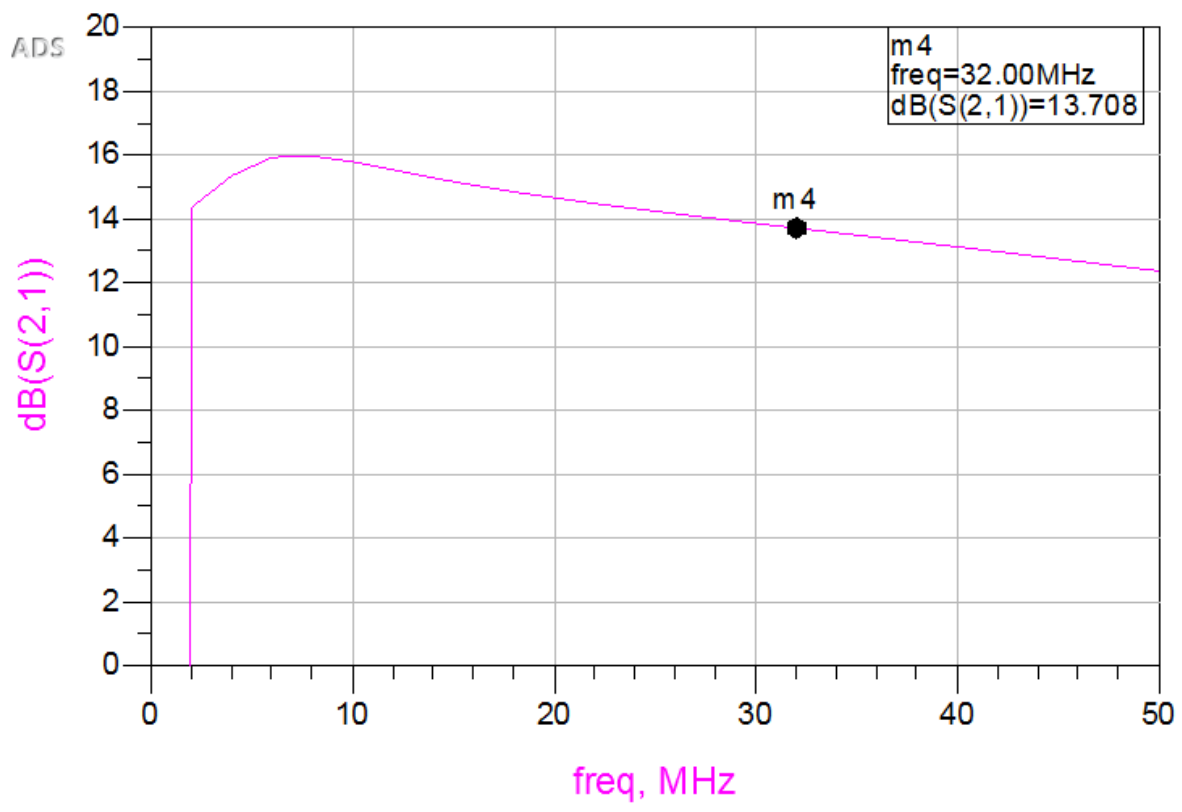

D)

Figure 7. Simulated magnitudes of S-parameters of the Cascode LNA. A) |S11|. B) |S12|. C) |S22|. D) |S21|

*E.* Linearity of the Cascode LNA

There are characteristics that are related with the non-linearity of an amplifier which are a third order intermodulation distortion and 1 dB compression point (P1dB) [7-8]. The P1dB point is the input power that causes the gain to decrease 1 dB from the normal expected linear gain. It is the point where the amplifier goes into compression and becomes non-linear. Therefore operation of the Cascode LNA should occur below this point in the linear region. In the most general sense, the output response of a nonlinear amplifier can be modeled as a Taylor series in terms of the input signal voltage that is given in equation (3) from equation (10.37) in [7]:

$$v_o = a_o + a_1\, v_i + a_2\, v_i^2 + + a_3\, v_i^3 + \cdots \qquad (3)$$

At the gain compression where a single-frequency sinusoid is given in equation (4) and it is applied to the input of a general nonlinear network, such as an amplifier, equation (10.39) in [7]:

$$v_i = V_o \cos \omega_o \mathrm{t} \qquad (4)$$

Equation (3) gives the output voltage as

$$v_o = a_o + a_1\, V_o \cos \omega_o \mathrm{t} + a_2\, V_o^2 \cos^2 \omega_o \mathrm{t} + a_3\, V_o^3 \cos^3 \omega_o \mathrm{t} + \cdots \qquad (5)$$

This result leads to the to the voltage gain of the signal component at frequency $\omega_o$:

$$G_v = \frac{v_o}{v_i} = a_1 + \frac{3}{4}\, a_3\, V_o^2 \qquad (6)$$

Where we have retained only terms through the third order.

The result of equation (6) shows that the voltage gain is equal to $a_1$, the coefficient of the linear term, as expected, but with an additional term proportional to the square of the input voltage amplitude. In most practical amplifiers $a_3$ typically has the opposite sign of $a_1$, so that the output of the amplifier tends to be reduced from the expected linear dependence for large values of $V_o$. This effect is called gain compression, or saturation.

For a third order intermodulation distortion, when an amplifier or other circuit becomes non-linear, it will begin to produce harmonics of the amplified inputs. The second, third, and higher harmonics are usually outside of the amplifier bandwidth, so they are usually easy to filter out if they are a problem. However, non-linearity will also produce a mixing effect of two or more signals. If the signals are close together in frequency, some of the sum and difference frequencies called intermodulation products produced can occur within the bandwidth of the amplifier. These cannot be filtered out, so they will ultimately become interfering signals to the main signals to be amplified. The third order intersection point, IP3 value is an imaginary point that indicates when the amplitude of the third-order products equals the input signals. This point is never reached, as the amplifier will saturate before this condition can occur. The IP3 point is typically about 10 dB above the 1-dB compression point. The simulation result of linearity analysis of the Cascode LNA is shown in Fig. 8.

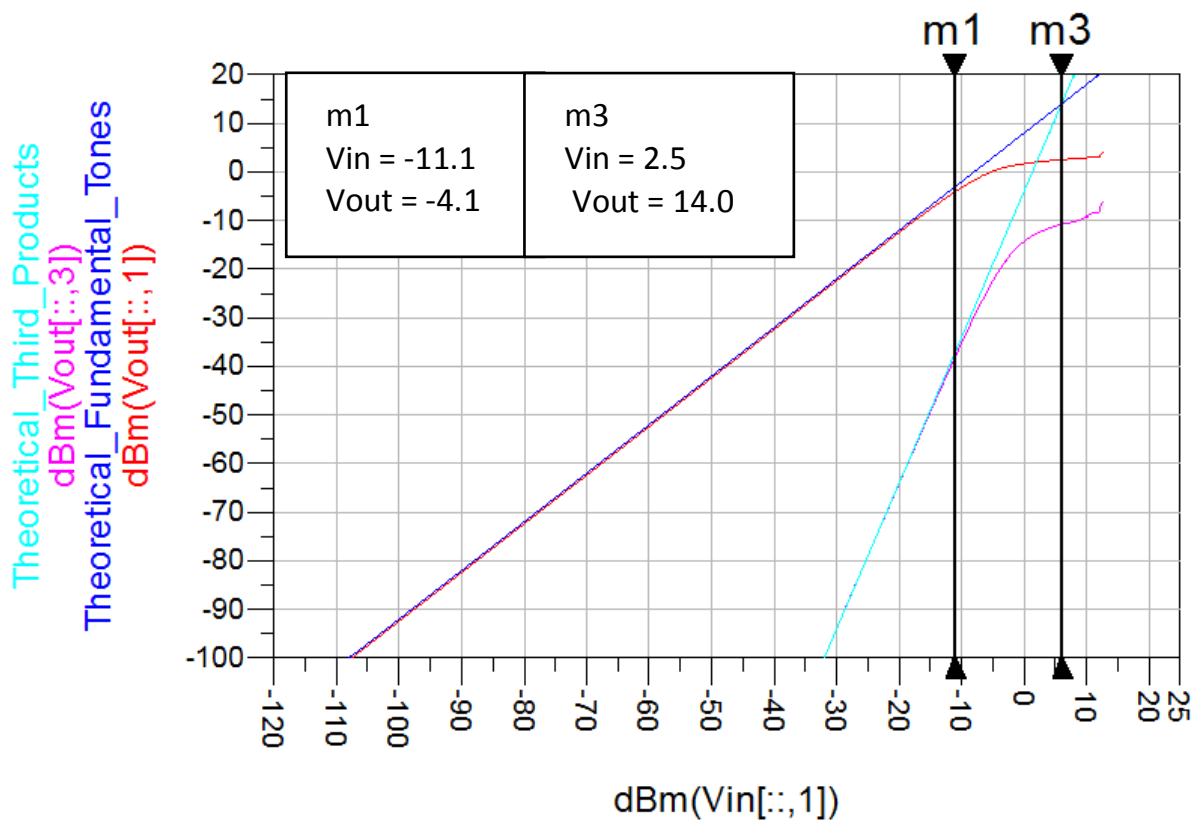

Figure 8. Simulated linearity analysis of the Cascode LNA. The parameter, Vin is the input power. The parameter, Vout is the output power

The figure above showed the gain compression of the Cascode LNA, P1dB = -11.1 dBm which is the input power that causes the gain to decrease about 1 dB from the normal expected linear gain. It is also showed the third order intersection point, IP3 = 14 dBm with reference to the output.

### III. Results

The PCB layout design of the Cascode LNA is fabricated and the measurements are performed while the Cascode LNA was inside a closed shielded box. The fabricated Cascode LNA is shown in Fig. 9.

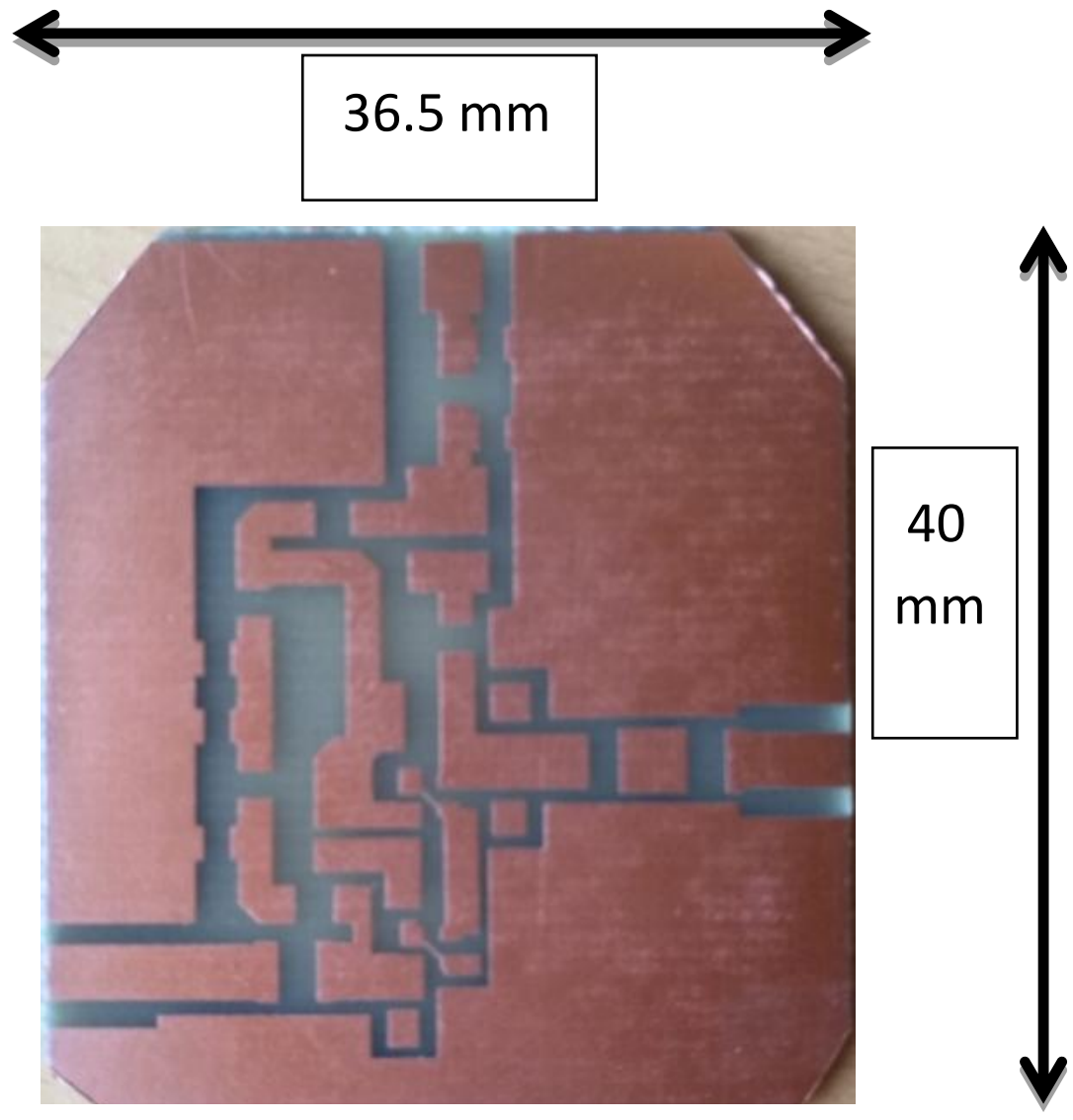

A)

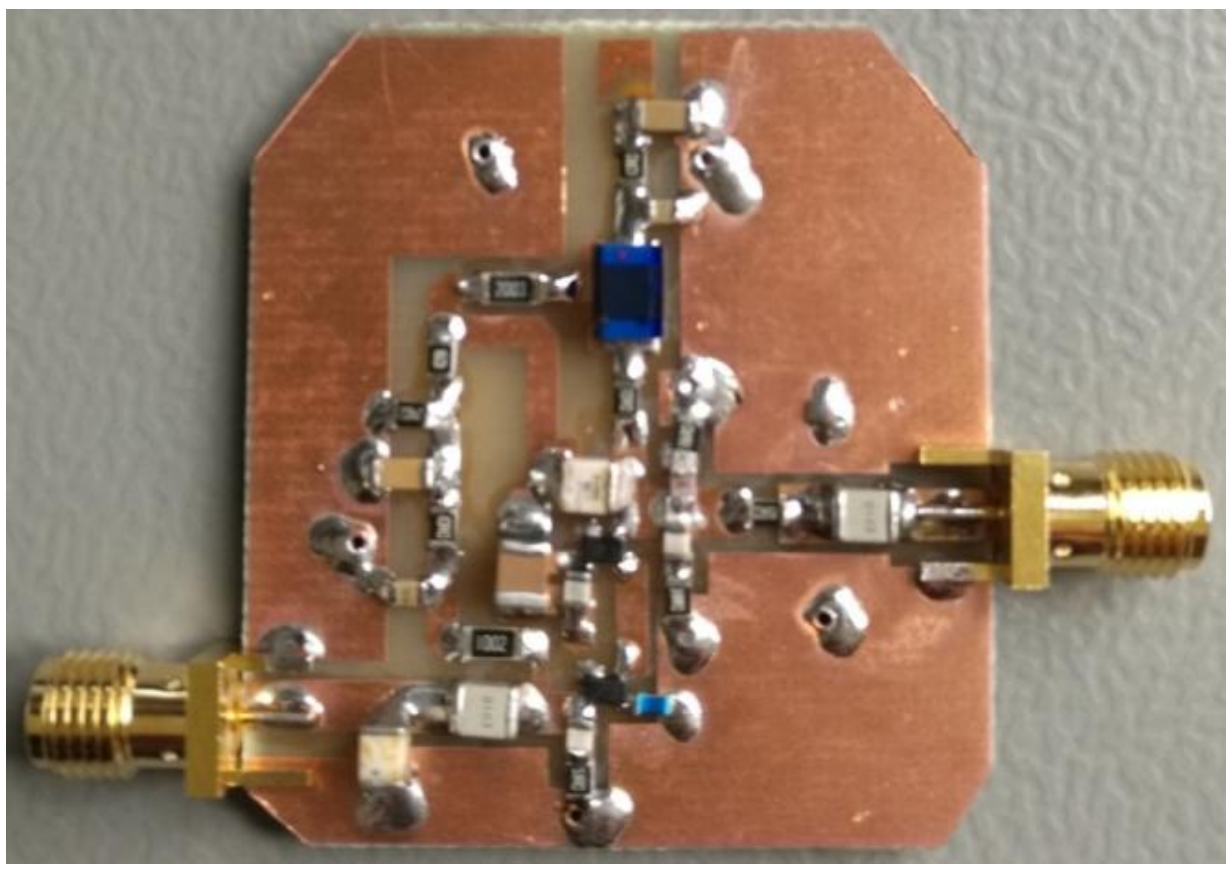

B)

Figure 9. The fabricated Cascode LNA. A) PCB of the Cascode LNA without the Components. B) PCB of the Cascode LNA and the components

The performance of the fabricated Cascode LNA needs to be measured and analyzed. Therefore it performed measurements with Vector Network Analyzer, Spectrum Analyzer, RF generator and Noise Source.

The measurements of the S-parameters are taken [18] as shown in Fig. 10. The measurements of the S-parameters at 32 MHz showed the gain of the Cascode LNA is 11.6 dB and the output return loss is 21.1 dB. Furthermore, the input return loss is 0.12 dB and the reverse isolation (Input-Output) is 63.8 dB. The Mu Stability Factor $\mu$ is plotted as shown in Fig. 11. The achieved measured factor $\mu$ is bigger than one over 6 GHz Band.

The analyzing of linearity of the fabricated Cascode LNA is achieved by conducting measurements with the Spectrum Analyzer and RF generator. It achieved 1 dB gain compression point by measuring the output power vs the swept input power and the results are plotted in Fig. 12.

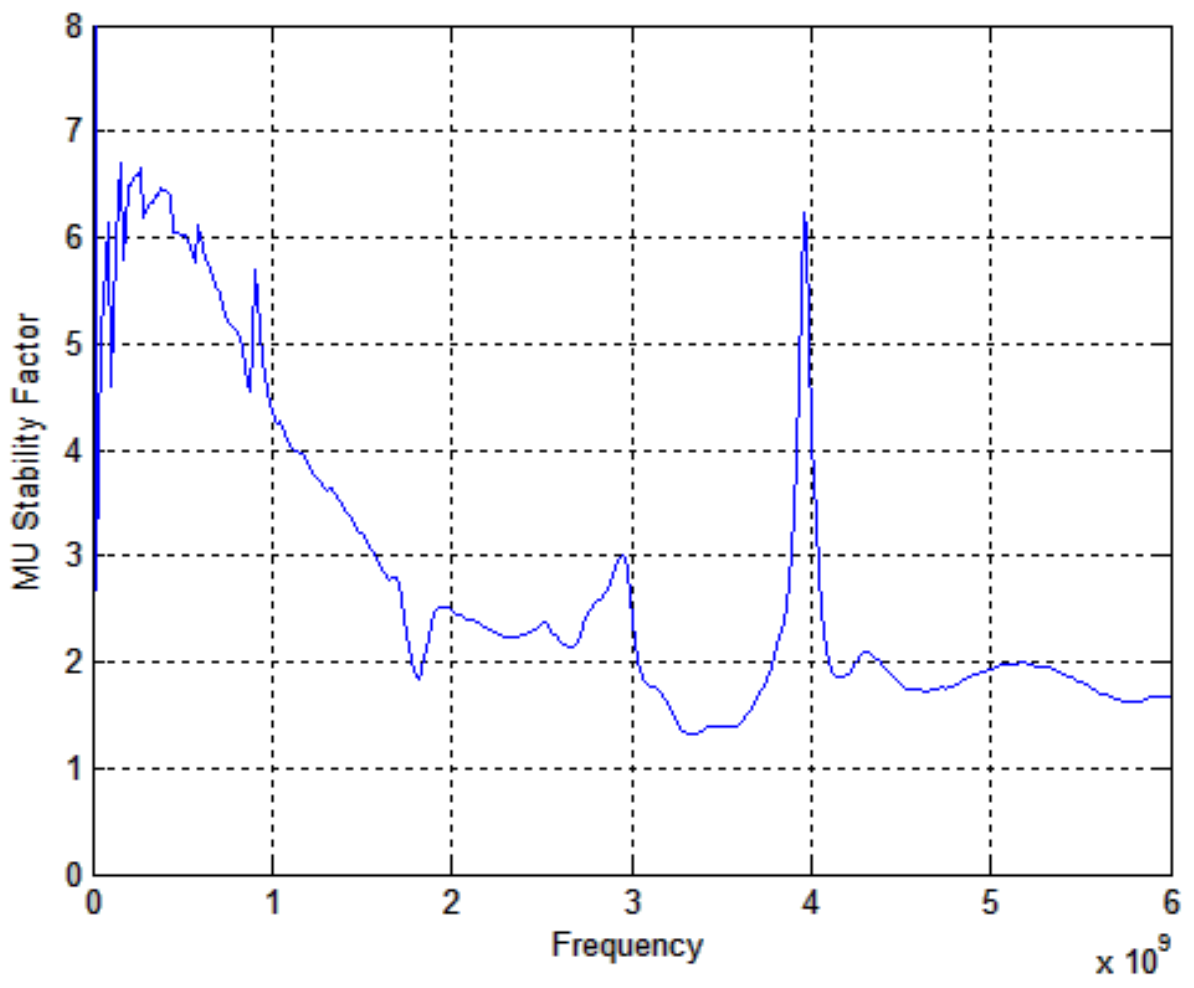

Figure 11. Measured Mu Stability Factor $\mu$ of the Cascode LNA

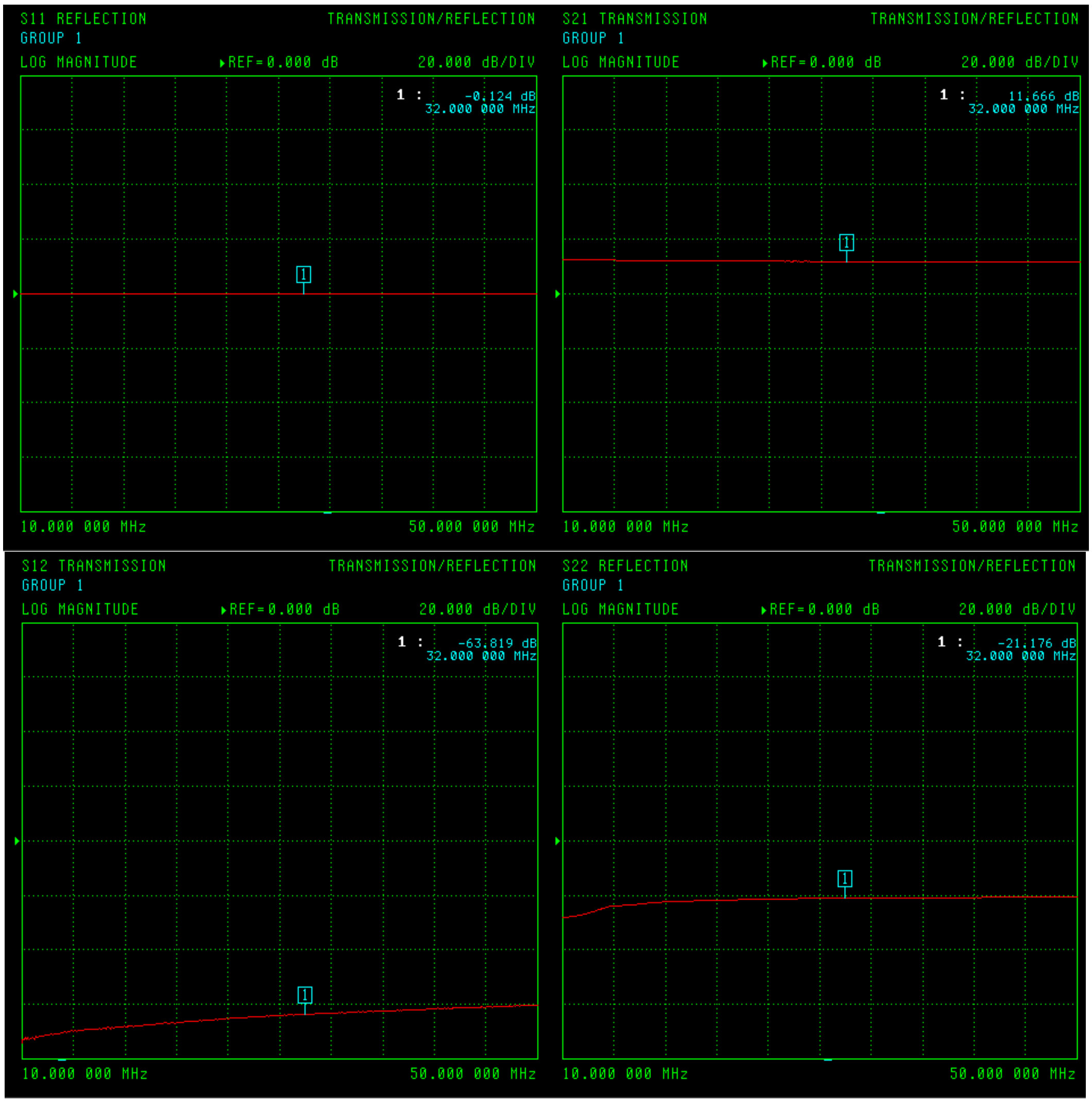

Figure 10. Measured S-parameters of the Cascode LNA

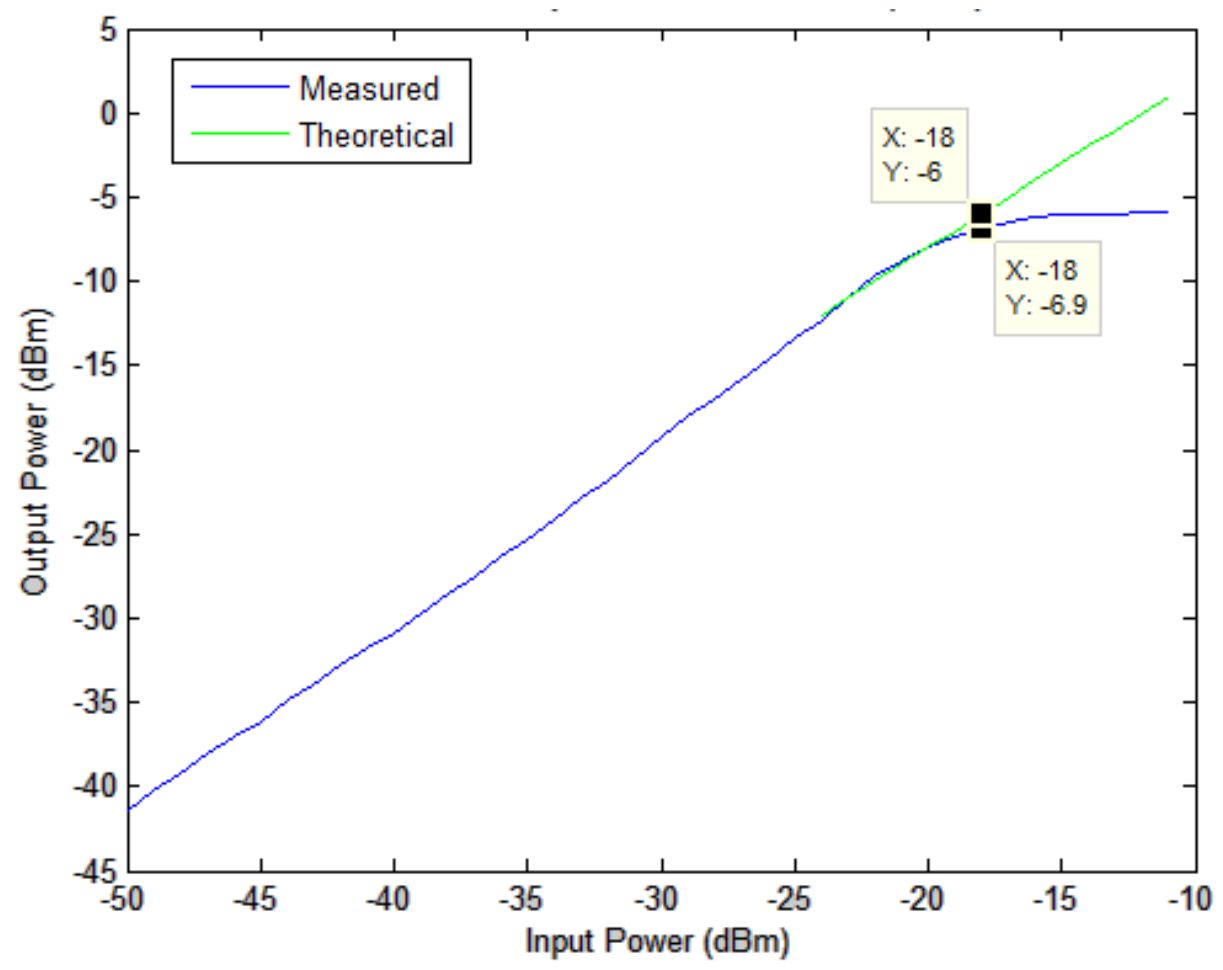

Figure 12. Measurements of input power vs output power at the Cascode LNA

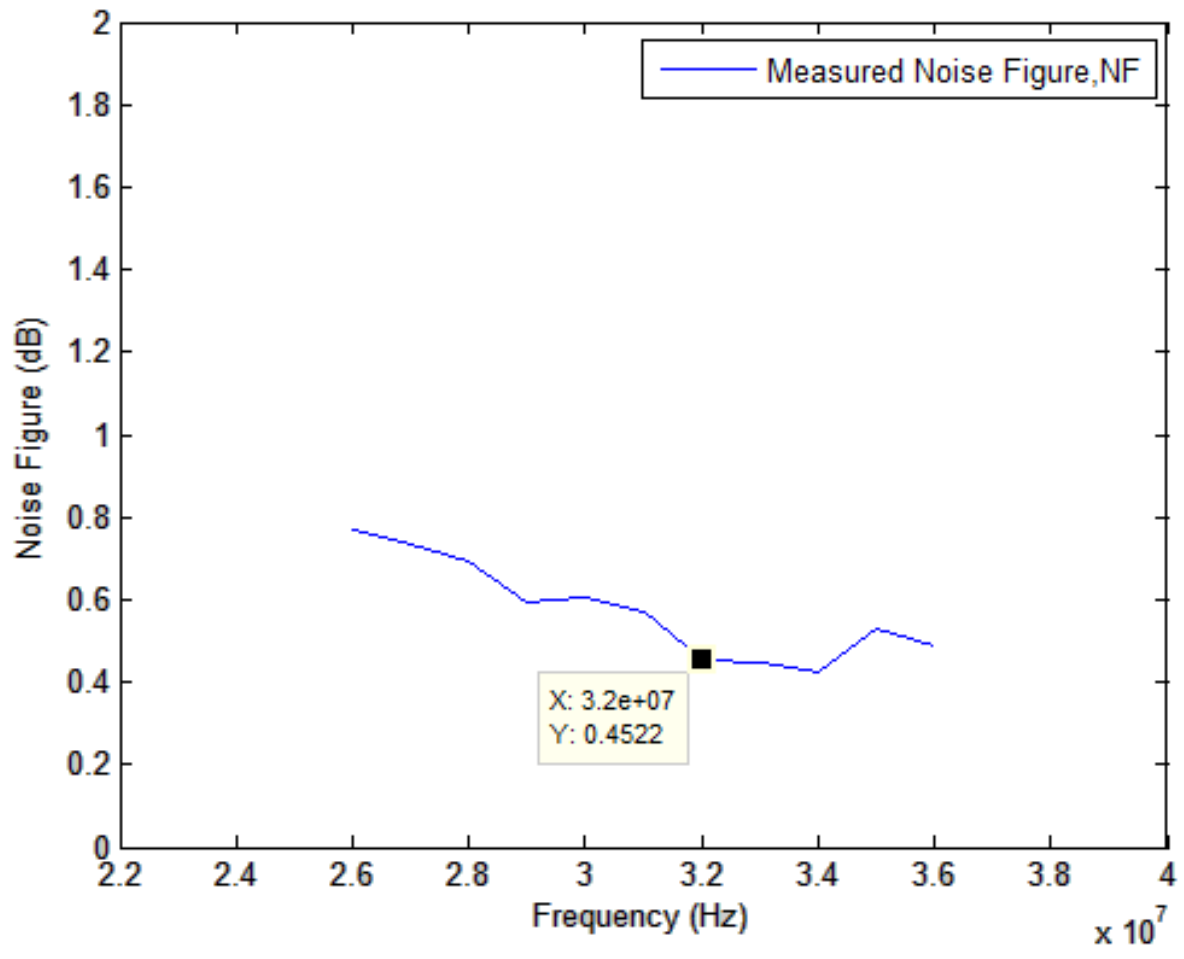

Figure 13. Measurements of noise figure of the Cascode LNA

The results of the measurements from the figure above showed the slope of the gain is nearly linear and the 1-dB compression point is about -18 dBm with respect to the input power. The measurement of the 1-dB compression point is about -6.9 dBm with respect to the output power. The gain compression point (P1dB) is found from the measurements above and it used to estimate the IP3 levels where the IP3 point is typically about 10 dB above the 1-dB compression point. Thus the estimate third order intercept point IP3 became -6.9 [dBm] + 10 [dBm] = 3.1 dBm with reference to the output.

The noise figure measurement of the Cascode LNA is conducted by using the Spectrum Analyzer. In order to measure noise figure of the Cascode LNA, a calibrated noise source is used at the measurement. The input return loss of the Cascode LNA is measured to 0.12 dB which it will imply inaccurate noise figure measurement. Therefore it used Attenuator at the noise figure measurement to remove the impact of input reflection of the Cascode LNA on the accuracy of the measured noise figure.

The spectrum analyzer will measure the total noise figure of the system which is the Attenuator and the Cascode LNA. In order to extract the only noise figure measurement of the Cascode LNA from the total noise figure measurement of the system, it applied Friis noise figure formula for cascaded devices [19]. The estimate measured noise figure of the Cascode LNA is shown in Fig. 13 where the estimate measured noise figure NF = 0.45 dB at 32 MHz. The measured noise figure of the Cascode LNA is conducted at room temperature T = 296.5 K [20].

The measurements on the fabricated Cascode LNA are done and the Table 1 will show summary of performance of the Cascode LNA at the measurements and the simulation that includes PCB layout with components parasitics.

| | Simulated Value | Measured Value |
|---|---|---|
| **Operation frequency** | 32 MHz | 32 MHz |
| **Gain** | 13.7 dB | 11.6 dB |
| **Noise figure at T= 296.5 K** | 0.38 dB | 0.45 dB |
| **Output return loss** | 24.1 dB | 21.1 dB |
| **Input return loss** | 0.21 dB | 0.12 dB |
| **Input-Output isolation** | 93 dB | 63.8 dB |
| **Output IP3** | 14 dBm | 3.1 dBm |
| **P1dB with respect to input** | -11.1 dBm | -18 dBm |
| **Stability** | Unconditional | Unconditional |
| **Source impedance** | 50 Ω | 50 Ω |
| **Load impedance** | 50 Ω | 50 Ω |
| **Supply voltage** | 5 volts | 5 volts |
| **Drain current** | 18 mA | 15 mA |

Table 1. Analyzing of performance of the Cascode LNA at the design implementing process

## IV. Conclusions

The design and developing of an Ultra-Low Noise Amplifier could achieve the goal of the research. The Cascode LNA had extremely good desired performance at the measurements. The operation frequency is 32

MHz and the estimate measured noise figure NF = 0.45 dB at source impedance 50 Ω is very low and this will improve the diagnostic accuracy of MRI system which is vitally important in medical applications. The Cascode LNA is unconditionally stable which is desirable to interface low impedance sources. The main objectives of the LNA design are obtained in the Cascode LNA performance as shown in Table 1 considering the low noise figure and unconditionally stability. The Cascode amplifier topology with Low Noise PHEMT transistor-ATF54143 could be successfully designed as an Ultra- Stable Low Noise Amplifier.

## Acknowledge

The research in this paper was prepared at Electromagnetic Systems group that is a part of Electrical Engineering department at Technical University of Denmark (DTU) in DK- Lyngby in fulfilment of the requirements for acquiring a MSc. in Electrical Engineering with Wireless Engineering Specialism. I want to thank Associate Professor Vitaliy Zhurbenko for his advising and supporting in this research.